\documentclass[sigconf,screen,nonacm,authorversion]{acmart}
\AtBeginDocument{%
  }

\acmISBN{978-1-4503-XXXX-X/2018/06}
\acmISBN{978-1-4503-XXXX-X/2018/06}
\usepackage{todonotes}
\usepackage{xcolor}
\usepackage{booktabs}
\usepackage{multirow}
\usepackage{longtable}
\usepackage[table]{xcolor}

\newcommand{\para}[1]{\vspace{0.3em}\noindent\textbf{#1}~}

\begin{document}

\title[Calibrating Student Trust in AI and Human Responses Through Mutual Theory of Mind]{\textit{Do We Know What They Know We Know?}\\Calibrating Student Trust in AI and Human Responses Through Mutual Theory of Mind}

\author{Olivia Pal}
\orcid{0009-0009-7489-4194}
\authornote{Both authors contributed equally.}
\affiliation{%
 \institution{University of Illinois Urbana-Champaign}
 \city{Urbana}
 \state{Illinois}
 \country{USA}}
 \email{opal2@illinois.edu}

\author{Veda Duddu}
\orcid{0009-0001-6443-6239}
\authornotemark[1]
\affiliation{%
  \institution{University of Illinois Urbana-Champaign}
 \city{Urbana}
 \state{Illinois}
 \country{USA}}
\email{vduddu2@illinois.edu}

\author{Agam Goyal}
\orcid{0009-0009-5989-2887}
\affiliation{%
  \institution{University of Illinois Urbana-Champaign}
 \city{Urbana}
 \state{Illinois}
 \country{USA}
}
\email{agamg2@illinois.edu}

\author{Drishti Goel}
\orcid{0009-0000-6713-9240}
\affiliation{%
  \institution{University of Illinois Urbana-Champaign}
 \city{Urbana}
 \state{Illinois}
 \country{USA}
}
\email{drishti4@illinois.edu}

\author{Koustuv Saha}
\orcid{0000-0002-8872-2934}
\affiliation{%
  \institution{University of Illinois Urbana-Champaign}
 \city{Urbana}
 \state{Illinois}
 \country{USA}
}
\email{ksaha2@illinois.edu}

\renewcommand{\shortauthors}{Pal and Duddu et al.}


\begin{abstract}
Trust and reliance are often treated as coupled constructs in human-AI interaction research, with the assumption that calibrating trust will lead to appropriate reliance. We challenge this assumption in educational contexts, where students increasingly turn to AI for learning support. Through semi-structured interviews with graduate students (N=8) comparing AI-generated and human-generated responses, we find a systematic dissociation: students exhibit high trust but low reliance on human experts due to social barriers (fear of judgment, help-seeking anxiety), while showing low trust but high reliance on AI systems due to social affordances (accessibility, anonymity, judgment-free interaction). Using Mutual Theory of Mind as an analytical lens, we demonstrate that trust is shaped by epistemic evaluations while reliance is driven by social factors---and these may operate independently. 
\end{abstract}



\maketitle
\section{Introduction and Background}
As AI systems become increasingly ubiquitous in educational tasks, students turn to generative AI tools like ChatGPT for problem-solving, explanation, and learning support~\cite{han2024teachers, park2024promise}. 
Yet, students' use of AI reveals a persistent tension: while some students exhibit over-reliance, accepting AI outputs without evaluation~\cite{abbas2024harmful}, others show algorithmic aversion, avoiding AI tools despite their potential benefits~\cite{shrivastava2025understanding}. 
These patterns suggest failures in how trust in AI is calibrated.


Importantly, the above behaviors exhibited by students cannot be explained by trust alone. 
Trust and reliance, though often used interchangeably, are distinct constructs. 
Trust, a psychological attitude, is defined as a willingness to be vulnerable based on positive expectations of another's behavior~\cite{mayer1995integrative}, encompassing beliefs about competence, benevolence, and integrity. 
In contrast, reliance is behavioral: the act of depending on a system's outputs when making decisions~\cite{lee2004trust}.
Recent work reveals that users frequently rely on AI outputs despite lacking genuine trust, particularly under time constraints or in difficult decisions~\cite{jacovi2021formalizing,nourani2019effects}. 
In such cases, reliance may be driven by perceived expertise gaps, cognitive load reduction, or opportunities to shift accountability for errors, rather than genuine confidence in system correctness.
As a result, reliance decisions can be strategically motivated and decoupled from trust attitudes.


This decoupling is reflected in prior findings on algorithmic aversion---where users reject algorithmic recommendations even when they outperform human judgment, and algorithmic appreciation, where users over-rely on AI uncritically~\cite{dietvorst2015algorithm,gaube2021ai}.
In fact, the same users may exhibit both patterns across contexts, suggesting that reliance is not only shaped by calibrated trust, but also by situational and affective factors.


Explainable AI (XAI) approaches have sought to address these challenges by promoting transparency as a mechanism for appropriate reliance~\cite{ribeiro2016should,kim2025fostering}.
However, responses do not necessarily lead to better calibration, and can also increase inappropriate reliance by fostering unwarranted confidence~\cite{vasconcelos2023explanations,bansal2021does,pareek2024effect, buccinca2021trust} while revealing output inconsistencies lowers perceived AI capacity despite improving comprehension \cite{lee2024one}.
Users with limited domain expertise often default to broad reliance regardless of accuracy, while experts exhibit more nuanced patterns---appropriately rejecting some flawed advice but also dismissing valid suggestions due to professional identity concerns or automation-related biases~\cite{zhang2020effect}. 
Moreover, the relationship between trust and reliance is strongly shaped by social context. 
For example, in legal and organizational contexts, reliance varies more with social accountability pressures than measured trust, and social proof often supersedes performance-based trust~\cite{green2019principles, yin2019understanding, parasuraman2010complacency}.

Despite these insights, prior work largely frames divergences between trust and reliance as calibration failures to be corrected~\cite{ribeiro2016should, zhang2020effect, bansal2021does}. 
What remains underexplored is whether, in socially rich contexts such as education, trust and reliance may be \textit{structurally} decoupled---following different motivational pathways shaped by social dynamics rather than epistemic evaluation alone \cite{ehsan2021expanding}.
Understanding this distinction is critical because interventions designed to increase trust (e.g., improving accuracy or explainability) may fail to change actual student behavior if reliance is primarily driven by social affordances such as accessibility, anonymity, or fear of judgment.
One promising way to interrogate this possibility is through users' mental representations of AI systems.

Theory of Mind (ToM)---the capacity to infer others' beliefs, intentions, and emotions---enables humans to interpret behavior and coordinate toward shared goals~\cite{baron2012, baron1985, frith&frith,wang2021towards}. Understanding how people assess others' mental states has been essential to explaining social decision-making and interpersonal trust. For example, prior research demonstrates that cooperation and trust depend on accurately representing another's mental perspective~\cite{baron1985, frith&frith, Gray2007}. Specifically, in human-AI interaction, ToM provides a critical lens to understand how humans perceive AI systems. Users frequently attribute mental states to AI despite the absence of genuine cognitive processing, thereby influencing trust, expectations, and perceived competence~\cite{Waytz2014,10.1145/2157689.2157696}. Language models exhibiting human-like behavior elicit anthropomorphic interpretations, often leading users to overestimate their capacities~\cite{Gray2007, Nass2000,wang2021towards}. 

Building on the above, recent work on Mutual Theory of Mind (MToM) emphasizes that humans and AI systems co-construct mental models through interaction~\cite{10.1145/2157689.2157696}. These mental models shape how users evaluate trustworthiness and make reliance decisions. When students attribute human-like understanding to AI, they may trust it yet rely on it inappropriately, while students perceiving AI as lacking understanding may distrust it yet rely on it for convenience. Human attributions often exceed actual system capacity~\cite{Liao2022,liao2022designing,kollerup2024can}, suggesting these mental models may influence trust and reliance independently.

Educational settings introduce additional complexities as students increasingly adopt generative AI tools (e.g., ChatGPT) for coursework support, from problem-solving and writing to brainstorming ideas, with roughly two-thirds of undergraduates using AI chatbots weekly~\cite{chan2023students, pitts2025understanding, pitts2025students,kim2024evallm}. Many students view AI as a convenient, personalized tutor that is easily accessible for immediate help, providing on-demand responses and feedback~\cite{chan2023students, sun2024would}. 
A the same time, researchers observe oscillations between \textit{overreliance}---accepting AI outputs without verification, risking shallow learning and reduced critical thinking~\cite{pitts2025students,li2023appropriate,gaube2021ai,duhaylungsod2023chatgpt}---and \textit{underreliance} driven by distrust or low confidence in AI systems~\cite{cheng2023overcoming,shao2025role}. 
Trust itself is highly context-dependent: greater statistical knowledge can increase trust in familiar situations while reducing trust in high-stakes contexts~\cite{marmolejo2025factors}


Motivated by these tensions, we explore how students' trust judgments translate into reliance on AI-assisted learning, and how social and goal-driven pressures can override epistemic evaluations.
We conducted a qualitative study in which students ($N$=8) compared AI-generated and human-generated responses under varying labeling conditions. 
This design enables us to observe how students form mental models of the explainer (AI vs. human), how those models shape trust, and when students nevertheless choose to rely (or not rely) on an explanation in their learning workflow. Concretely, we ask: \textbf{Does trust predict reliance in AI-assisted learning contexts?}



Our findings reveal a systematic dissociation: students exhibit high trust but low reliance on human experts due to social barriers (e.g., fear of judgment, anxiety seeking-help), while simultaneously showing low trust but high reliance on AI systems, primarily due to accessibility and anonymity. This suggests that trust and reliance may operate independently in educational contexts, with trust driven by epistemic evaluations and reliance driven by social affordances.
\section{Methods}

\begin{table}[t]
\footnotesize
\sffamily
\centering
\caption{\textbf{Example Stimulus.} Each question is paired with four explanation variants: correct and incorrect versions from both human and AI sources. See \autoref{tab:all-stimuli} in \autoref{app:appendix} for the complete stimulus set.}
\label{tab:example-stimulus}
\begin{tabular}{lp{0.25\columnwidth}p{0.5\columnwidth}}
\arrayrulecolor{black}
\rowcolor{gray!20} 
\textbf{Condition} & \textbf{Human-Generated} & \textbf{AI-Generated (RAG)} \\
\midrule
\rowcolor{blue!10} 
\multicolumn{3}{c}{\textit{Question: What is loop current?}} \\
\midrule
\rowcolor{green!5} 
\textbf{Correct} & 
Loop current is the imagined current flowing around a loop. & 
Loop current is the assumed current circulating around a mesh in mesh analysis, with a defined direction, used to solve circuit equations. \\
\midrule
\rowcolor{red!5} 
\textbf{Incorrect} & 
Loop current is the imagined current flowing through a single component & 
Loop current is the branch current; substitution swaps loop for branch.\\
\bottomrule
\end{tabular}
\small
\end{table}


For our study, we recruited eight graduate students in Computer Science (ages 18-32, five female, three male) via class forums and referrals. See \autoref{tab:participants} in \autoref{app:appendix} for details on our participant pool.
All involvement was voluntary, anonymous, and adhered to institutional ethical standards. 
Each participant engaged in a 60-75 minute semi-structured interview that included pairwise comparison tasks centered on electrical engineering problems, such as the functioning of diodes. 
Each comparison featured one explanation produced by a human and one generated by AI, emulating real-world scenarios in which students must determine whether to believe AI or human sources. 
We included both factually accurate and inaccurate responses to examine how participants evaluate and reason about responses. Following established approaches for studying AI perceptions \cite{lee2024one}, we measured trust and reliance through comparative judgments and stated preferences, enabling investigation of the mechanisms underlying these decisions.

We employed a within-subjects design~\cite{greenwald1976within} to monitor the evolution of trust judgments across successive exposures to source information. 
The sessions were conducted in three phases: \textbf{(1) Baseline (Blinded):} Participants assessed explanation pairs devoid of source labels, establishing content-based assessment baselines; \textbf{(2) Construction (First Labeled):} Random assignment to \emph{Revealed} (accurate AI/Human labels) or \emph{Mislabeled} (swapped labels) conditions, investigating initial mental model formation that integrates source labels with content quality; \textbf{(3) Revision (Second Labeled):} Exposure to the alternative labeling condition, assessing mental model resilience and adaptability. This counterbalanced design differentiated the cognitive processes of construction and revision within individuals. 

Following each comparison, participants identified their preferred explanation and assessed their confidence on a scale of 0 to 10. Decision time and consistency checks (using a few repeated question pairs) were used to assess hesitation, reliability, and stability in their trust choices. Participants employed think-aloud techniques, verbalizing their reasoning during assessments, yielding substantial qualitative data.

\noindent{\textbf{Dataset Creation:}}
Realistic quality variation in both human and AI responses was necessary to facilitate pairwise comparison. 
Both human experts and AI systems lack perfect accuracy; human responses may arise from misunderstandings, whilst AI responses might include hallucinations and erroneous generalizations~\cite{dror2011paradox, xu2024hallucination}. 
Our dataset comprised responses from both experts and non-experts.

\noindent\textbf{Domain Selection:} The principles of electrical engineering established ideal circumstances for examining trust: participants required adequate knowledge to interact with the material, while their partial understanding necessitated dependence on other resources. Personal dependence on expert counsel is dependent on metacognitive evaluation of one's knowledge in relation to the perceived competence of the advisor~\cite{meshi2012expert}. STEM graduate students generally have basic knowledge but lack expertise---an ``intermediate knowledge zone'' prone to overconfidence, making it suitable for examining trust in external counsel~\cite{sanchez2024intermediate}.

\noindent\textbf{Human Expert Responses:} Accurate responses obtained from \href{https://www.khanacademy.org/science/electrical-engineering}{Khan Academy} and \href{https://spinningnumbers.org/}{Spinning Numbers}. Erroneous responses systematically integrated recorded student misconceptions~\cite{kuccukozer2007secondary,cohen1983potential}, akin to memory research paradigms employing semantically related lures to investigate false recall~\cite{roediger1995creating}, thereby guaranteeing authentic conceptual errors rather than inadequate writing.

\noindent\textbf{AI Responses:} RAG~\cite{lewis2020retrieval} anchored AI responses in the same source material as human responses, guaranteeing content equivalence and minimizing hallucination. We instructed LLMs to generate accurate responses while intentionally incorporating established misconceptions. This guaranteed that all stimuli derived from the identical basic material for systematic comparison.
The dataset has 8 subject areas, each containing 5 to 7 questions. Every question comprises four options: Human Correct (H-C), Human Incorrect (H-IC), AI Correct (AI-C), AI Incorrect (AI-IC).

\para{Qualitative Analysis.}
We employed reflexive thematic analysis~\cite{braun2019reflecting} due to its adaptability in integrating both theory-driven and data-driven insights. While addressing our research question about whether trust predicts reliance in AI-assisted learning contexts, we remained receptive to emergent themes from participant experiences.

All eight interviews were audio-recorded and transcribed verbatim (approximately 7.5 hours), maintaining speech expressions, pauses, and non-verbal signals, enhancing contextual depth. Transcripts were anonymized using participant identifiers (P1-P8).
We utilized hybrid coding that integrates both deductive and inductive methodologies. Our preliminary codebook encompassed theoretical constructs (trust, dependence, mental models, source attribution, comprehension) while being receptive to unforeseen patterns. Two co-first authors independently analyzed all transcripts 
assigning descriptive codes to data segments that highlighted significant aspects or innovative discoveries, while recording emergent patterns, tensions, and initial interpretations.

Following independent coding, co-first authors assembled for three collaborative refining sessions: (1) Comparing codes across transcripts, addressing discrepancies, clarifying definitions, consolidating redundant codes, subdividing broad codes, and incorporating emergent codes not initially captured; (2-3) Organizing codes into candidate themes by discerning meaning patterns throughout the dataset, highlighting interview trajectories to uncover latent themes. Themes were polished to effectively address research questions while capturing complexity and nuance.

\noindent\textbf{Validation:} Conclusive themes emerged through iterative analysis and confirmation with interview data. Each theme was corroborated by numerous extracts from participants, establishing robust patterns rather than isolated occurrences. We analyzed anomalous cases---responses that oppose prevailing patterns---to encompass the complete range of experiences. Four co-authors participated in ongoing discussions, ensuring reflexivity regarding how our AI experiences could influence interpretations, hence enhancing trustworthiness.

\section{Findings}
Our analysis of the think-aloud protocols and semi-structured interviews revealed that trust and reliance operate as distinct constructs when evaluating AI-generated versus human-generated responses. Rather than moving in tandem as dominant frameworks predict~\cite{lee2004trust}, these constructs dissociate systematically, driven by different factors. 

\para{Trust-Reliance Dissociation}
Participants constructed fundamentally different mental models of human and AI sources that produced systematic mis-calibration in opposite directions. 
 




\begin{figure*}
    \centering
    \includegraphics[width=0.85\linewidth]{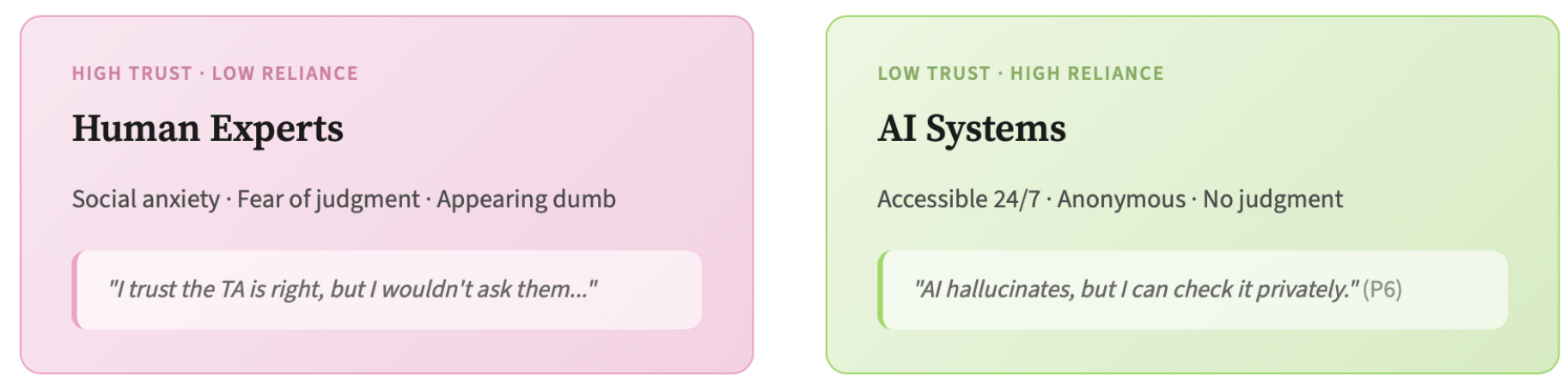}
    \caption{\textbf{Trust-Reliance Dissociation.} Students exhibit an inverse relationship between trust and reliance when choosing between human experts and AI systems. Despite trusting human experts to provide accurate information, students often avoid seeking their help due to social anxiety and fear of judgment. Conversely, students readily rely on AI systems---despite potential hallucinations---because they offer 24/7 accessibility, anonymity, and a judgment-free environment where mistakes can be made privately.}
    \label{fig:dissociation}
\end{figure*}

Participants constructed fundamentally different mental models of human and AI sources that produced systematic miscalibration in opposite directions. Students consistently attributed strong epistemic credibility to human teaching assistants and instructors, citing instructional experience and reliability as reasons for trusting human experts. As P2 noted, \textit{"TAs have taught this before, they know where students struggle"}. However, actual reliance on these experts was paradoxically low, with participants reporting that they avoid human expertise due to social barriers such as fear of judgment, appearing incompetent, and anticipated embarrassment. 

Conversely, students expressed explicit skepticism about AI competence, citing hallucinations, shallow reasoning, and mathematical weakness as reasons for low trust. Students emphasized accessibility, anonymity, and lack of social cost as drivers of use, with P6 explaining: \textit{"I know AI hallucinates and makes mistakes, but I can check it privately without anyone judging me, so I use it way more than I probably should."}



\para{Why They Dissociate: Different Factors Shape Trust vs Reliance.}
Analysis revealed that trust and reliance are shaped by fundamentally different factors, allowing us to explain why they seem to operate independently. \textbf{Trust} is driven by epistemic evaluations, such as perceived competence, domain expertise, and reliability, while \textbf{Reliance} is driven by social and contextual affordances---accessibility, anonymity, and social costs. These factors operate orthogonally, with our participants scoring human experts higher on epistemic factors but low on social affordances, producing high trust and low reliance. AI systems on the other hand, score lower on epistemic factors but high on social affordances---producing low trust but high reliance. 

Participants P2 and P8 articulated this disconnect: P2 trusted human expertise \textit{("TAs have domain knowledge")} while avoiding consultation \textit{("I don't want to bother them")}, whereas P8 distrusted AI accuracy \textit{("It's terrible at math")} while relying on it \textit{("If I am stuck at 2 AM, AI is my only option")}:

This mechanism explains the dissociation: when different factors shape trust versus reliance, the two constructs decouple. The interventions that increase AI trustworthiness (accuracy, explainability) may not reduce overreliance if the primary drivers are accessibility and anonymity. Similarly, interventions that increase trust in human experts will not increase reliance if social barriers remain unaddressed. 


\section{Discussion, Implications, and Future Directions}
Our findings suggest design directions for educational AI systems that address trust and reliance as distinct targets. We hope to explore these implications by testing new systems that may have the following design features.  
We find that trust and reliance operate through distinct pathways in education, with trust driven by epistemic evaluations and reliance by social affordances. This has implications for intervention design. Current approaches focus on epistemic calibration---expressing uncertainty \cite{kim2025fostering}, revealing inconsistencies \cite{kim2024m}, providing sources and varied explanation types and deploying trustworthiness signals \cite{kollerup2024can}. These shape trust but not reliance when students use AI for accessibility and anonymity rather than correctness. Students may distrust AI yet rely on it for anonymous, accessible help. Conversely, trustworthiness cues may boost confidence in experts without reducing help-seeking anxiety. This gap separates what systems optimize (trust) from what drives behavior (social affordances) \cite{ehsan2023charting}.

Educational AI must pair epistemic features with reliance management. Uncertainty expressions should redirect to human help, inconsistency revelations should normalize confusion and suggest TA consultation, and responses should contain gaps prompting human consultation. Therefore, systems could adapt based on student mental models that detect when repeated use signals social barriers and respond by reducing accessibility or scaffolding consultation \cite{wang2021towards}. Beyond individual interactions, institutions must address structural barriers: anonymous channels, normalized struggles, and low-stakes formats that reduce fear of judgment.

\subsubsection{Design AI as a Bridge to Human Help, Not a Replacement}

The pivotal implication of our findings is that future design must tackle the decoupling between trust and reliance in AI use. Students' overuse of AI is driven less by high levels of trust and more by accessibility and anonymity. Therefore, instead of positioning AI as a replacement for human instruction, it should be designed as low-stakes scaffolds that bridge students to human consultation and support.

\para{Nudge collaborative problem-solving offline.} 
After repeated interactions on the same concept, an AI system could encourage students to transition from AI assistance to human consultation. The system could generate specific questions students might ask their TA or instructor based on their interaction history. 
Also, it could provide responses with intentional gaps, where discussing the missing pieces with a classmate or TA would help students connect the ideas.


\noindent\textbf{Reduce social friction of help-seeking} Students avoid human experts due to fear of judgment and beliefs that their questions are invalid or embarrassing. AI can address these barriers by normalizing concerns---for instance, noting that a majority of students struggle with the same concept—and providing anonymized pathways to human consultation, such as reformulating questions for forum posts. The system could validate common confusions explicitly and generate specific question starters for office hours, reducing the friction of initiating help-seeking.


\subsubsection{Design AI for Appropriate Trust, Not Overconfidence}
While bridging to human help addresses over-reliance, students still need AI systems they can appropriately trust. Our findings revealed that transparency features---particularly citations, reasoning traces---can override inherent source preferences and enable better trust calibration.

\noindent\textbf{Enable verification through citations.} Students expressed that citation availability would reverse their hesitancy to trust AI over human experts. Systems should link every factual claim to verifiable source material when possible and display where information came from, noting which sources cover which specific concepts. When sources conflict or are unavailable, the system should flag uncertainty, presenting multiple perspectives rather than a single answer.


\noindent\textbf{Scaffold comprehension before expecting trust judgments.} 
Students noted that they struggle to trust answers they cannot comprehend. Systems should check prerequisite knowledge before presenting complex responses, asking whether students need foundational concepts explained first. Comprehension checks---such as asking students to explain key concepts back---can help the system gauge whether explanations succeeded. Additionally, systems can employ progressive complexity, starting simple and increasing detail only when students demonstrate understanding.

\bibliographystyle{ACM-Reference-Format}
\bibliography{references.bib}

\appendix
\section{Appendix}\label{app:appendix}

\begin{table}[h]
\centering
\caption{Participant Demographics (N=8)}
\label{tab:participants}
\begin{tabular}{cccc}
\hline
\textbf{ID} & \textbf{Age} & \textbf{Gender} & \textbf{Academic Level} \\
\hline
P1 & 18-25 & M & Graduate \\
P2 & 26-32 & M & Graduate \\
P3 & 18-25 & F & Graduate \\
P4 & 18-25 & M & Graduate \\
P5 & 18-25 & F & Graduate \\
P6 & 18-25 & M & Graduate \\
P7 & 18-25 & F & Graduate \\
P8 & 18-25 & F & Graduate \\
\hline
\end{tabular}
\end{table}

\onecolumn
\begin{longtable}{p{0.15\columnwidth}lp{0.3\columnwidth}p{0.35\columnwidth}}
\caption{Dataset Overview: Representative Question-Response Pairs}
\label{tab:all-stimuli} \\
\arrayrulecolor{black}
\rowcolor{gray!20}
\textbf{Question} & \textbf{Condition} & \textbf{Human-Generated} & \textbf{AI-Generated (RAG)} \\
\midrule
\endfirsthead

\multicolumn{4}{c}%
{{\bfseries \tablename\ \thetable{} -- continued from previous page}} \\
\rowcolor{gray!20}
\textbf{Question} & \textbf{Condition} & \textbf{Human-Generated} & \textbf{AI-Generated (RAG)} \\
\midrule
\endhead

\midrule
\multicolumn{4}{r}{{Continued on next page}} \\
\endfoot

\bottomrule
\endlastfoot

\rowcolor{green!5}
\multirow{2}{*}{\parbox{0.15\columnwidth}{What is Kirchhoff's Law?}} & Correct &
Kirchoff's Law is the sum of all currents flowing into a node equals the sum of currents flowing out of the node & 
Kirchhoff's laws: KCL—sum of currents at a node is zero; KVL—sum of voltages around a closed loop is zero. \\
\cmidrule(lr){2-4}
\rowcolor{red!5}
& Incorrect & 
Kirchoff's Law is the sum of all voltages flowing into a node equals the sum of the voltages flowing out of the node & 
Replace voltages with currents; the sum of currents around a loop equals zero.\\
\midrule

\rowcolor{green!5}
\multirow{2}{*}{\parbox{0.15\columnwidth}{What is node voltage?}} & Correct & 
Node voltage is the potential difference between two nodes of a circuit & 
Node voltage is the voltage difference between a given node and another point in the circuit. \\
\cmidrule(lr){2-4}
\rowcolor{red!5}
& Incorrect & 
Node voltage is the sum of voltages of two nodes of a circuit. & 
Node voltage is the total of the voltages at two nodes in the circuit.\\
\midrule

\rowcolor{green!5}
\multirow{2}{*}{\parbox{0.15\columnwidth}{What is the reference node?}} & Correct & 
One of the nodes is selected to be a reference node. All node voltages are measured with respect to this one reference node. & 
One node is chosen as the reference, serving as the zero-potential or ground point from which all node voltages are measured. It does not affect current flow but provides a common voltage reference for the circuit. \\
\cmidrule(lr){2-4}
\rowcolor{red!5}
& Incorrect & 
One of the nodes is selected to be a reference node, which changes the current flow in the circuit. All node voltages are measured with respect to this reference node. & 
One node is chosen as the reference, altering the current flow in the circuit. By designating two such zero-potential references, all measured voltages effectively double.\\
\midrule

\rowcolor{green!5}
\multirow{2}{*}{\parbox{0.15\columnwidth}{What is a loop?}} & Correct & 
A loop is any closed path around a circuit. To trace a loop, one starts at the component terminal and traces a path through connected elements until one gets back to the starting point. & 
A loop is any closed path in a circuit where current can circulate, returning to the starting point. \\
\cmidrule(lr){2-4}
\rowcolor{red!5}
& Incorrect & 
A loop is any open path around a circuit. To trace a loop, one starts at a component terminal and traces a path through connected elements without returning to the starting point & 
A loop is any path in a circuit that does not return to the starting point.\\
\midrule

\rowcolor{green!5}
\multirow{2}{*}{\parbox{0.15\columnwidth}{What is a mesh?}} & Correct & 
A mesh is a restricted kind of loop, and a mesh is a loop that contains no other loops. & 
A mesh is a closed loop in a planar circuit that does not enclose any other loops. \\
\cmidrule(lr){2-4}
\rowcolor{red!5}
& Incorrect & 
A mesh is a restricted kind of loop, and a mesh is a loop that contains at least one other loop inside it & 
A mesh is not a closed loop in a circuit.\\
\midrule

\rowcolor{green!5}
\multirow{2}{*}{\parbox{0.15\columnwidth}{What is loop current?}} & Correct & 
Loop current is the imagined current flowing around a loop. & 
Loop current is the assumed current circulating around a mesh in mesh analysis, with a defined direction, used to solve circuit equations. \\
\cmidrule(lr){2-4}
\rowcolor{red!5}
& Incorrect & 
Loop current is the imagined current flowing through a single component & 
Loop current is the branch current; substitution swaps loop for branch.\\
\midrule

\rowcolor{green!5}
\multirow{2}{*}{\parbox{0.15\columnwidth}{What is a semi-conductor?}} & Correct & 
Semiconductor materials fall between insulators and conductors. They usually act like insulators, but we can control how much they conduct by changing the way they are made and by applying voltages. & 
A semiconductor is a solid material whose conductivity lies between conductors and insulators and can be controlled by doping, temperature, and electric fields. \\
\cmidrule(lr){2-4}
\rowcolor{red!5}
& Incorrect & 
Semiconductor materials fall between insulators and conductors. They usually act like conductors, but we can control how much they insulate by changing the way they are made and by applying voltages & 
A semiconductor is a perfect conductor; conductivity is unchanged by temperature, doping, or electric fields.\\
\midrule

\rowcolor{green!5}
\multirow{2}{*}{\parbox{0.15\columnwidth}{What does bias mean?}} & Correct & 
Bias in the sense of pulling towards or favoring one side. Forward bias means an externally applied voltage is tugging the diode towards the forward conducting side of its i-v curve. Reverse bias is the opposite, a negative external voltage pulls the diode into its reverse bias region. & 
Bias is a DC voltage or current applied to set the operating point of an active device (e.g., transistor). \\
\cmidrule(lr){2-4}
\rowcolor{red!5}
& Incorrect & 
Bias in the sense of pulling towards or favoring one side. Forward bias means an externally applied voltage is tugging the diode towards the backward conducting side of its i-v curve. Reverse bias is the opposite, a negative external voltage pulls the diode into its reverse bias region. & 
Bias is a DC voltage or current applied to regulate the device's temperature.\\
\midrule

\rowcolor{green!5}
\multirow{2}{*}{\parbox{0.15\columnwidth}{How to increase LED brightness?}} & Correct & 
One way to get more diode current is to reduce the resistance. Lower resistance increases the current produced for a given voltage. This leads the LED to become brighter. & 
Lower the series resistor value or raise supply voltage within LED ratings; or use a proper driver (transistor/MOSFET) to increase current. \\
\cmidrule(lr){2-4}
\rowcolor{red!5}
& Incorrect & 
One way to get more diode current is to increase the resistance. Higher resistance increases the current produced for a given voltage. This leads the LED to become brighter. & 
Increasing current actually dims the LED.\\
\midrule

\rowcolor{green!5}
\multirow{2}{*}{\parbox{0.15\columnwidth}{Remove resistor for max brightness?}} & Correct & 
If the resistance becomes smaller and smaller, current required at a given voltage will be significantly higher. This leads to either the diode burning out from excess heat or the diode survives, the battery runs down in just a little while. & 
If the resistance keeps decreasing, the current at a fixed voltage rises sharply, which can overheat or burn out the diode, or quickly drain the battery. Bad idea — without a resistor, current runs away and damages the LED, so a current-limiting device is essential. \\
\cmidrule(lr){2-4}
\rowcolor{red!5}
& Incorrect & 
If the resistance becomes smaller and smaller, current required at a given voltage will be significantly lower. This means the diode battery can run for much longer. & 
If the resistance keeps decreasing, the current at a fixed voltage becomes lower, allowing the diode to operate longer on the same battery. In this case, current is self-limiting and brightness remains safe.\\
\midrule

\rowcolor{green!5}
\multirow{2}{*}{\parbox{0.15\columnwidth}{What is a photodiode?}} & Correct & 
This diode has a window to let light fall directly on the silicon surface. The current in the diode is proportional to the intensity of light. & 
A photodiode is a PN junction semiconductor device with a window that lets light fall directly on the silicon surface, generating a current proportional to the light's intensity. It functions as a light detector. \\
\cmidrule(lr){2-4}
\rowcolor{red!5}
& Incorrect & 
This diode has a window to let light fall directly on the silicon surface. The current in the diode is inversely proportional to the intensity of light. & 
A photodiode is a PN junction semiconductor device with a window that lets light fall directly on the silicon surface, generating a current inversely proportional to the light's intensity. It is not a light detector.\\

\end{longtable}
\twocolumn

\end{document}